# Multimodal pattern formation in phenotype distributions of sexual populations

Michael Doebeli[1], Hendrik J. Blok[1], Olof Leimar[2] & Ulf Dieckmann[3,4]

[1] Department of Zoology and Department of Mathematics, University of British Columbia, 6270 University Boulevard, Vancouver B.C. Canada, V6T 1Z4
[2] Department of Zoology, University of Stockholm, SE-106 91 Stockholm, Sweden
[3] Evolution and Ecology Program, International Institute for Applied Systems Analysis, Schlossplatz 1, A-2361 Laxenburg, Austria
[4] Section Theoretical Biology, Institute of Biology, Leiden University, Kaiserstraat 63, NL-2311 GP Leiden, The Netherlands

September 25, 2006

**ABSTRACT**

During bouts of evolutionary diversification, such as adaptive radiations, the emerging species cluster around different locations in phenotype space, How such multimodal patterns in phenotype space can emerge from a single ancestral species is a fundamental question in biology. Frequency-dependent competition is one potential mechanism for such pattern formation, as has previously been shown in models based on the theory of adaptive dynamics. Here we demonstrate that also in models similar to those used in quantitative genetics, phenotype distributions can split into multiple modes under the force of frequency-dependent competition. In sexual populations, this requires assortative mating, and we show that the multimodal splitting of initially unimodal distributions occurs over a range of assortment parameters. In addition, assortative mating can be favoured evolutionarily even if it incurs costs, because it provides a means of alleviating the effects of frequency dependence. Our results reveal that models at both ends of the spectrum between essentially monomorphic (adaptive dynamics) and fully polymorphic (quantitative genetics) yield similar results. This underscores that frequency-dependent selection is a strong agent of pattern formation in phenotype distributions, potentially resulting in adaptive speciation.

# 1. INTRODUCTION

Explaining the origin of diversity is a core problem in evolutionary biology that continues to receive much attention from both empiricists and theoreticians (Coyne & Orr 2004; Dieckmann *et al.* 2004). The process of diversification can be described as evolutionary change taking place in phenotype space. If individual organisms are assessed for their phenotypes, populations can be represented by the corresponding phenotype distributions, giving information about their abundance in the population. A single, ancestral population would typically yield a unimodal phenotype distribution, with the average phenotype being at or close to the distribution's peak. Processes of speciation can then often be described as the splitting of an ancestral and unimodal phenotype distribution into two (or more) peaks or modes, so that the descendent species emerging from speciation correspond to different peaks of the phenotype distribution. On the phenotypic level, speciation can thus cause pattern formation: during speciation, unimodal phenotype distributions may become multimodal.

Traditional explanations for such pattern formation through speciation are based on geographic isolation: different, but phenotypically similar, subpopulations of an ancestral species come to occupy different and mutually isolated habitats, in which they embark on different evolutionary trajectories. These trajectories may eventually take the populations evolving in different habitats to different locations in phenotypes space, so that the joint phenotype distribution of all descendent species becomes multimodal. It is important to appreciate that this allopatric mode of phenotypic pattern formation and speciation results from geographical isolation, rather than from ecological interactions within the ancestral population.

The situation is reversed for sympatric processes of speciation, which unfold due to ecological interactions within the ancestral population, rather than as a consequence of geographical isolation. For example, when phenotypes differ in their resource preference, and when most individuals in an ancestral population prefer similar resources, selection may favour rare phenotypes with a different resource preference. In this case, diversification of the ancestral population may be an adaptive response to the detrimental effects of frequency-dependent competition. In general, adaptive speciation occurs when an ancestral lineage splits into phenotypically diverging descendent lineages due to disruptive selection caused by frequency-dependent interactions (Dieckmann *et al.* 2004). In this mode of speciation, pattern formation in phenotype space is caused by interactions that are intrinsic to the ancestral population. The theoretical framework of adaptive dynamics predicts that such adaptive diversification can occur under a wide variety of ecological scenarios (Metz *et al.* 1996; Geritz *et al.* 1998; Dieckmann & Doebeli 1999; Doebeli & Dieckmann 2000, 2003; Dieckmann *et al.* 2004; Kisdi & Gyllenberg 2005). In this framework, adaptive diversification is epitomized by the phenomenon of evolutionary branching, which occurs when frequency-dependent selection drives a population towards a point in phenotype space at which selection turns disruptive. Evolutionary branching can be characterized mathematically and is a generic outcome of adaptive dynamics models (Metz *et al.* 1992, 1996; Geritz *et al.* 1998; Kisdi & Gyllenberg 2005).

Most models of evolutionary branching are based on a number of seemingly significant simplifying assumptions. Chief among those are the assumptions that reproduction is asexual, and that populations are essentially monomorphic at all times (except when branching occurs, after which each of the emerging lineages is assumed to be essentially monomorphic). Obviously, both of these assumptions are often violated in real populations. It is thus important that it has been shown that evolutionary branching is also a robust outcome in asexual models of polymorphic populations (Metz *et al.* 1996; Meszéna *et al.* 2005), and that a number of recent models have incorporated explicit genetics to study adaptive speciation in sexual populations (Doebeli 1996; Dieckmann & Doebeli 1999; Drossel & McKane 2000; Dieckmann *et al.* 2004; Kondrashov & Kondrashov 1999; Kisdi & Geritz 1999; Geritz & Kisdi 2000; Doebeli & Dieckmann 2003; Doebeli 2005; Schneider & Bürger 2006; Bürger & Schneider 2006; Bürger *et al.* 2006). In sexual populations under disruptive selection, random mating typically prevents speciation, so that diversification requires the presence of assortative-mating mechanisms ensuring that individuals preferentially mate with similar phenotypes. Such mechanisms have been considered in models with genetic architectures based on small to intermediate numbers of loci with additive effects (see articles cited above). One general conclusion of such studies is that adaptive speciation, or pattern formation in phenotype space, is possible in sexual populations when mating is assortative.

It has recently been suggested by Polechová & Barton (2005) that occurrences of adaptive speciation in sexual populations could often be a consequence of the particular genetic models used, and that other genetic models would not generate diversification in sexual populations even with assortative mating. One reason for this caveat might be that in genetically explicit models with a finite number of loci and with finite allelic effects, a population's variance is automatically constrained, leading to more intense intraspecific competition and thus strengthening disruptive selection. In models with more flexible genetic architectures, intraspecific competition might simply result in increased population variance. In particular, for populations described by a continuous phenotype distribution (rather than by a single monomorphic type or by the frequencies of a finite number of types), one might have the intuitive expectation that frequency-dependent competition merely flattens unimodal phenotype distributions, thus compensating for the effects of competition. According to this intuition, frequency-dependent selection would not result in pattern formation in phenotype space, i.e., in a bimodal or multimodal split of the phenotype dis-tribution, and hence would not result in adaptive speciation.

To describe the dynamics of continuous phenotype distributions under frequency-dependent competition, Polechová & Barton (2005) used the "infinitesimal" model of quantitative genetics, which assumes a large (infinite) number of unlinked loci with additive effects (Roughgarden 1979; Bulmer 1980). Polechová & Barton (2005) claim that in such models, frequency dependence never leads to adaptive speciation even if mating is assortative. This would support the intuitive notion that frequency dependence can generate increased population variance, but not pattern formation in the form of bimodal or multimodal phenotype distributions. These questions are very interesting and deserve further study. In this paper we use a more general class of models to show that frequency-dependent competition in sexual populations indeed leads to pattern formation in phenotype space under many circumstances.

The intuitive notion that in models for continuous phenotype distributions, frequency dependence only leads to increased variance, but not to phenotypic clusters, thus turns out to be wrong in general. Instead, if mating is assortative, frequency-dependent competition often generates multiple phenotypic modes also in infinitesimal models. Since a population's split reduces the strength of disruptive selection, assortative mating facilitates the evolutionary response to frequency dependence. Consequently, there is selection for assortative mating in initially randomly mating populations, in which segregation and recombination would otherwise prevent the emergence of multiple modes. This is why pattern formation in phenotype space is a possible outcome of frequency-dependent competition in infinitesimal models of sexual populations.

Our results show that, with regard to adaptive diversification, the outcomes of asexual adaptive dynamics models at one end of the spectrum, and of infinitesimal sexual models at the other end, are surprisingly congruent. In the sexual models, assortative mating is required for adaptive speciation to occur, but in both types of model the emergence of distinct phenotypic clusters out of unimodal or even monomorphic ancestral populations can readily be caused by frequency-dependent ecological interactions. This pattern formation in sexual models could be an important mechanism underlying the instability and disruption of the sexual continuum of phenotypes (Maynard Smith & Szathmáry 1995; Noest 1997), and hence could help address the fundamental question why life forms appear to cluster phenotypically (Coyne & Orr 2004).

## 2. MODEL DESCRIPTION

Below we introduce the dynamics of the density distribution $\phi(x)$ of a quantitative character $x$ in a sexual population.

**(a)** *Ecological dynamics*

The ecological model underlying our analysis is an extension of Lotka-Volterra competition equations to polymorphic populations, in which the competitive impact of phenotype $y$ on a phenotype $x$ is measured by the competition kernel $\alpha(x-y)$. For a focal phenotype $x$, the total competitive impact experienced in a population described by the distribution $\phi$ is given by the convolution

$$(\alpha * \phi)(x) = \int \alpha(x-y)\phi(y)dy. \tag{1}$$

In the asexual case, the dynamics of the distribution $\phi$ are then given by the following partial differential equation,

$$\frac{\partial \phi}{\partial t} = r\phi\left(1 - \frac{\alpha * \phi}{K}\right) = r\phi - r\phi \cdot \alpha * \phi / K. \tag{2}$$

Here $r$ is the intrinsic growth rate, which we assume to be independent of the phenotype $x$, and $K(x)$ determines the carrying capacity as a function of $x$. Thus $-r\phi \cdot \alpha * \phi/K$ corresponds to the usual competition term in Lotka-Volterra models, whereas $r\phi$ describes exponential population growth.

**(b)** *Mating and reproduction*

We incorporate sexual reproduction following standard procedures (Roughgarden 1979; Bulmer 1980; see also Polechová & Barton 2005). We assume that matings are initiated bilaterally. The probability of mating between two phenotypes $u$ and $v$ is therefore proportional to the product of two preference functions, which we assume to be Gaussian,

$$A(u,v) = \frac{1}{\sqrt{2\pi}\sigma_A}\exp\left(-\frac{(u-v)^2}{2\sigma_A^2}\right) \cdot \frac{1}{\sqrt{2\pi}\sigma_A}\exp\left(-\frac{(u-v)^2}{2\sigma_A^2}\right), \tag{3}$$

where $\sigma_A$ is a measure for the degree of assortment: large $\sigma_A$ correspond to random mating, while small $\sigma_A$ correspond to assortative mating (note that we will always assume here that assortative mating occurs with respect to the quantitative character that determines the ecological interactions).

In accordance with the assumptions underlying the infinitesimal model of quantitative genetics (Bulmer 1980), we assume that a mating between phenotypes $u$ and $v$ produces a Gaussian offspring distribution $N_{(u+v)/2,\sigma_f}(x)$, with a mean equalling the midparent value $(u+v)/2$ and a variance of $\sigma_f^2$.

To establish a baseline case, we assume that all phenotypes have the same per capita birth rate. This means that the relative contribution a mating with phenotype $v$ makes to the offspring pool of a given phenotype $u$ must be normalized by the total amount of mating that phenotype $u$ participates in,

$$N(u) = \int A(u,v)\phi(v)dv. \tag{4}$$

Then the distribution of offspring with phenotypes $x$ produced by phenotype $u$ is given by

$$\frac{1}{N(u)}\int \phi(v)A(u,v)N_{(u+v)/2,\sigma_f}(x)dv. \tag{5}$$

Finally, the total density of offspring at phenotype $x$ resulting from all possible matings is given by

$$\beta(x) = \int \phi(u)\left[\frac{1}{N(u)}\int \phi(v)A(u,v)N_{(u+v)/2,\sigma_f}(x)dv\right]du. \tag{6}$$

The mating scheme just described for the infinitesimal model is a direct extension of the one used in Dieckmann & Doebeli (1999) for genetically explicit multilocus models.

Putting everything together, we obtain the following equation for the dynamics of phenotype distributions in sexual populations,

$$\frac{\partial \phi}{\partial t} = r\beta - r\phi \cdot \alpha * \phi / K. \tag{7}$$

The essential parameters in this dynamical system are $\sigma_A$ (degree of assortment) and $\sigma_f$ (width of the so-called segregation kernel (Roughgarden 1979) that describes the offspring distribution of a given mating pair), as well as the functional forms of the ecological functions $\alpha$ and $K$. For numerical simulations of the partial differential equation (7), we always used carrying capacity functions $K$ with finite variance, which implies that phenotypes that are far from the optimal phenotype are not viable. This allows the numerical simulations to be restricted to a finite interval without creating artefacts.

**(c)** *Competition kernel and carrying capacity function*

It is already very interesting to study the dynamics of the asexual model, eq. (2), which is determined by the functions $\alpha$ and $K$. In particular, one can ask whether, for given functions $\alpha$ and $K$, equilibrium distributions of the asexual model exhibit phenotypic clustering in the form of multiple modes. For example, if the competition kernel $\alpha$ and the carrying capacity $K$ are both of Gaussian type with variances $\sigma_\alpha^2$ and $\sigma_K^2$, respectively, then the model has an equilibrium density distribution that is also Gaussian, with variance $\max(0, \sigma_K^2 - \sigma_\alpha^2)$ (if $\sigma_K^2 - \sigma_\alpha^2$ is negative, the equilibrium distribution has all its density concentrated at the maximum of $K$). In particular, with Gaussian $\alpha$ and $K$, equilibrium distributions of the asexual model never exhibit more than one phenotypic cluster.

It is known, however, that the asexual model with Gaussian ecological functions is structurally unstable (Sasaki and Ellner 1995; Sasaki 1997), and that generic choices for the ecological functions often lead to pattern formation with distinct phenotypic clusters (Meszéna *et al.* 2005). We therefore use competition kernels of the form

$$\alpha(x-y) = \exp\left(-\frac{|x-y|^{2+\varepsilon_\alpha}}{2\sigma_\alpha^{2+\varepsilon_\alpha}}\right) \tag{8}$$

and carrying capacity functions of the form

$$K(x) = K_0 \exp\left(-\frac{x^{2+\varepsilon_K}}{2\sigma_K^{2+\varepsilon_K}}\right). \tag{9}$$

Here the shape parameters $\varepsilon_\alpha$ and $\varepsilon_K$ measure deviations from the Gaussian case.

### (d) *Equilibrium distributions*

For $\varepsilon_\alpha = \varepsilon_K = 2$ (the "quartic" case in which the competition kernel and the carrying capacity are both platykurtic), it can easily be shown numerically that equilibrium distributions of the asexual model (2) have multiple peaks whenever $\sigma_\alpha$ is small enough.

By contrast, for the sexual model (7) with Gaussian ecological functions $\alpha$ and $K$ with variances $\sigma_\alpha^2$ and $\sigma_K^2$, one can show, by carrying out the various integrals introduced above, that a Gaussian equilibrium distribution exists whose variance $\sigma_{eq}^2$ satisfies the following equation,

$$\frac{2(\sigma_A^2 + \sigma_{eq}^2)^2}{4\sigma_{eq}^4(\sigma_f^2 + \sigma_{eq}^2) + 2\sigma_A^4(2\sigma_f^2 + \sigma_{eq}^2) + \sigma_A^2\sigma_{eq}^2(8\sigma_f^2 + 5\sigma_{eq}^2)}$$
$$= \frac{1}{2\sigma_{eq}^2} + \frac{1}{2(\sigma_\alpha^2 + \sigma_{eq}^2)} - \frac{1}{2\sigma_K^2}. \tag{10}$$

For example, in the case of random mating, $\sigma_A = \infty$, the variance of the Gaussian equilibrium distribution satisfies

$$\frac{1}{2\sigma_f^2 + \sigma_{eq}^2} = \frac{1}{2\sigma_{eq}^2} + \frac{1}{2(\sigma_\alpha^2 + \sigma_{eq}^2)} - \frac{1}{2\sigma_K^2}. \tag{11}$$

Similarly, in the case of extreme assortative mating, $\sigma_A = 0$, there is a Gaussian equilibrium distribution with a variance satisfying

$$\frac{1}{2(\sigma_f^2 + \sigma_{eq}^2)} = \frac{1}{2\sigma_{eq}^2} + \frac{1}{2(\sigma_\alpha^2 + \sigma_{eq}^2)} - \frac{1}{2\sigma_K^2}. \tag{12}$$

The existence of Gaussian equilibrium distributions in infinitesimal models in which the eco-logical functions have Gaussian form may be perceived as supporting the claim that frequency-dependent competition in polymorphic populations does not usually generate multimodal phenotype distributions. However, two important caveats need to be kept in mind. First, even though a Gaussian equilibrium distribution exists, it may not be stable under the dynamics given by eq. (7). Second, the existence of the Gaussian equilibrium given by eq. (10) depends on the assumption that the ecological functions $\alpha$ and $K$ have Gaussian form. As mentioned above, the asexual model with Gaussian ecological functions is structurally unstable, and hence there is no reason to believe that sexual models with non-Gaussian ecological functions and assortative mating would generally admit unimodal equilibrium distributions.

The use of Gaussian functions for $\alpha$ and $K$ has a long tradition in the literature (Roughgarden 1979). Unfortunately, other than for the fact that a Gaussian decrease in competitive effects and in carrying capacities appears to be heuristically appealing, there is no reason for using these particular functional forms. In fact, Ackermann & Doebeli (2004) have shown that the case in which both the competition kernel and the carrying capacity are Gaussian with finite variance cannot be derived from the underlying mechanistic consumer-resource model introduced by MacArthur (1972), which lies at the basis of most competition models for continuous characters (Roughgarden 1979). This in itself does not mean that the Gaussian case is biologically implausible, but it means that there is no biological reason why this case should receive preferential treatment over other, more general functions, such as those given by eqs. (8) and (9). In fact, the mathematical simplicity of the Gaussian case, which sometimes allows analytical equilibrium solutions, may lead to an undesirable bias towards drawing conclusions from a structurally unstable scenario (Meszéna *et al.* 2005). More general models, such as those based on eqs. (8) and (9), will generally yield more robust results, even though one typically has to resort to numerical simulations for solving the corresponding dynamical equations for the phenotype distribution.

## 3. RESULTS

Before we turn our attention to the effects of assortative mating on the dynamics of phenotype distributions in sexual populations, we mention two general conditions that are necessary for pattern formation to result in multimodal distributions. First, the width of the offspring distribution of a given mating pair, $\sigma_f$, must be small enough compared to the width of the carrying capacity function, $\sigma_K$. Wide offspring distributions tend to homogenize populations and hence to prevent pattern formation. Second, the force of frequency-dependent selection needs to be strong enough for the emergence of multiple phenotypic clusters. For our purposes, this means that in the ecological functions given by eqs. (8) and (9) the width of the competition kernel, $\sigma_\alpha$, must be small enough compared to the width of the carrying capacity function, $\sigma_K$. Wide competition kernels weaken frequency-dependent disruptive selection and hence prevent pattern formation.

**(a)** *Implications of assortative mating*

Even with these necessary conditions being satisfied, we never observed phenotypic pattern formation when mating was random, in which case the equilibrium distributions were invariably unimodal. However, strikingly different outcomes resulted when mating was assortative, i.e., for small enough $\sigma_A$. This is illustrated in figure 1, which shows stable equilibrium distributions of the infinitesimal model for different values of $\sigma_A$ for the case in which the competition kernel and the carrying capacity are both Gaussian. As we pointed out in the previous section, this model admits Gaussian equilibrium distributions with variances given by eq. (10). These equilibrium distributions are stable for high $\sigma_A$ (random mating, figure 1a) as well as for very low $\sigma_A$ (very strong assortment, figure 1d). In these cases, the numerical simulations are in exact agreement with the analytical predictions for the variances of the equilibrium distribution given by eq. (11) for $\sigma_A = \infty$ and by eq. (12) for $\sigma_A = 0$.

However, there is a range of intermediate values of $\sigma_A$ for which the Gaussian equilibrium distributions are unstable, and instead the dynamics converges to an equilibrium distribution exhibiting distinct phenotypic modes, as shown in figures 1b,c. Because mating is assortative, the phenotypic clusters emerging through such pattern formation represent incipient species: the resultant clusters are reproductively isolated to a large degree, with little gene flow occurring between them. To illustrate the niche partitioning between the incipient species, the grey lines in figure 1 show the carrying capacity function $K$, indicating the total available niche space. For figures 1b,c, the initial phenotype distributions were chosen to be very close to the Gaussian equilibrium distribution, but, rather than approaching this Gaussian equilibrium, the system diverges from these unimodal distributions and exhibits pattern formation. Our numerical simulations indicate that when the multimodal equilibrium distributions are stable, they are attractors for a large range of initial conditions. This is illustrated in figure 2 for the case shown in figure 1b.

We note that the fact that the Gaussian equilibrium is stable for very small $\sigma_A$ (figure 1d) is a consequence of the special and non-robust characteristics of the Gaussian case for the asexual model, in which Gaussian ecological functions always generate unimodal solutions (see previous section): for very strong assortative mating, the sexual model becomes similar to the asexual model (albeit even in the limit of $\sigma_A = 0$ the sexual model is not exactly equivalent to the asexual model unless $\sigma_f = 0$).

Figures 3a-d show examples of equilibrium distributions for quartic ecological functions, i.e., for $\varepsilon_\alpha = \varepsilon_K = 2$ in eqs. (8) and (9). Again, random mating results in unimodality (figure 3a), but assortative mating readily results in multimodal phenotype distributions (figures 3b-d). In this case, diversification occurs even for very strong assortative mating (figure 3d), corresponding to the fact that models with quartic ecological functions admit multimodal solutions even in the asexual case. In contrast to the case of Gaussian ecological functions, the existence of unimodal equilibrium distributions (stable or unstable) cannot be asserted when ecological functions are non-Gaussian. Even if such equilibrium distributions exist in the quartic case, our simulations indicate that they are never stable when assortment is strong enough. In particular, for the values of $\sigma_A$ used for figures 3a-d, the dynamics converge to the shown multimodal equilibrium distributions, independently of the various initial conditions that we tested.

In the quartic case, our extensive numerical simulations indicate that the dependence of pattern formation on the various parameters can be roughly summarized as follows. First, for multimodal pattern formation we have the basic requirement that $\sigma_\alpha$ must be small enough to produce frequency-dependent disruptive selection, i.e., $\sigma_\alpha < \sigma_K$. Second, both $\sigma_f$ and $\sigma_A$ need to be small compared to $\sigma_\alpha$ and $\sigma_K$. We have found that this can be approximately summarized by the two conditions $\sigma_f + \sigma_A < \sigma_\alpha$ and $\sigma_f + \sigma_A < \sigma_K/3$. Our simulations indicate that these conditions generally imply pattern formation in the quartic case. These conditions also apply in the case of Gaussian ecological functions, except that with Gaussian functions, we have the additional condition $\sigma_f < \sigma_A$. If this condition is not satisfied, the sexual system behaves like the Gaussian asexual model and possesses a stable unimodal distribution (figure 1d). On theoretical grounds, it is difficult to assess the biological relevance of the above conditions. There is at least some empirical support for the ecological condition $\sigma_\alpha < \sigma_K$ (Bolnick *et al.* 2003), and situations in which the genetic kernels (described by $\sigma_f$ and $\sigma_A$) are narrower than the ecological kernels (described by $\sigma_\alpha$ and $\sigma_K$) do not appear to be unrealistic.

Figure 4 further illustrates the generality of the phenomenon of diversification through pattern formation in phenotype space in the presence of assortative mating. In figure 4a, we considered different forms of the carrying capacity function by varying the shape parameter $\varepsilon_K$, while assuming a Gaussian form for the competition kernel ($\varepsilon_\alpha = 0$). For a given carrying capacity function $K$, we varied the assortative mating parameter $\sigma_A$ from values corresponding to random mating (right) to values representing strong assortment (left). For each parameter combination $(\sigma_A, \varepsilon_K)$, the figure indicates whether the resulting equilibrium phenotype distribution had a single or multiple modes. Analogously, in figure 4b we considered different forms of the competition kernel $\alpha$ by varying the shape parameter $\varepsilon_\alpha$, while assuming a Gaussian form for the carrying capacity function ($\varepsilon_K = 0$).

To produce figure 4, we used uniform initial phenotype distributions to start the dynamics for each tested parameter combination. However, the results were virtually identical when Gaussian initial distributions with unit variance were used. That these very different initial conditions yielded the same results underscores that the long-term dynamics of the models considered is largely independent of the initial conditions. Thus figure 4 shows that diversification resulting in multimodal phenotype distribution occurs for a wide range of assortative mating parameters, and for general classes of competition kernels and carrying capacity functions.

**(b)** *Evolution of assortative mating*

Given that assortative mating can facilitate phenotypic diversification due to frequency-dependent interactions, as evidenced in figures 1 and 2, it is natural to ask whether there is selection pressure on assortment itself to evolve in initially randomly mating populations. We analyze the selection acting on assortment in two steps. We first assume that the degree of assortment is asexually inherited (one could think of it as being maternally inherited), which permits an adaptive dynamics analysis. We then implement the sexual inheritance of the assortment trait based on standard quantitative genetics in an individual-based model.

For the adaptive dynamic analysis, we extended eqs. (7) to two types differing in their degree of assortment. This allows us to follow the dynamics of the phenotype distributions of the two different types, and in particular to determine when one type can invade the other. With $\phi_1(x)$ and $\phi_2(x)$ denoting the phenotype distributions of the two types with assortative mating parameters $\sigma_{A_1}$ and $\sigma_{A_2}$, respectively, the resulting dynamics are given by

$$\frac{\partial \phi_1}{\partial t} = r\beta_1 - r\phi_1 \cdot \alpha * (\phi_1 + \phi_2)/K, \tag{13}$$

$$\frac{\partial \phi_2}{\partial t} = r\beta_2 - r\phi_2 \cdot \alpha * (\phi_1 + \phi_2)/K. \tag{14}$$

Because the two types are ecologically equivalent, their per capita death rates $r \cdot \alpha * (\phi_1 + \phi_2)/K$ are equal, while their birth rates $\beta_1(x)$ and $\beta_2(x)$ may differ as a result of differential assortment. These birth rates are derived in the Appendix.

To understand the evolutionary dynamics of assortative mating, we used eqs. (13) and (14) to generate pairwise invasibility plots (Metz *et al.* 1996; Geritz *et al.* 1998). These are two-dimensional plots in which possible resident phenotypes are shown on the horizontal axis and possible mutant phenotypes on the vertical axis. For each resident-mutant pair $(\sigma_{A,\mathrm{res}}, \sigma_{A,\mathrm{mut}})$, we first let a population consisting only of the resident type reach equilibrium, and then introduced a mutant type at small total density, in order to evaluate whether the mutant's growth rate was positive or negative. The mutant's initial phenotype distribution was assumed to have the same shape as the resident's equilibrium distribution, but with a much reduced total density. Using eqs. (13) and (14), the mutant's growth rate was measured as the change in total density over a number of subsequent generations. This procedure generates a partitioning of the pairwise invasibility plot into plus-regions, indicating that for such resident-mutant pairs the mutant can increase when rare and hence will potentially invade the resident, and minus-regions, indicating that the mutant can-not invade the corresponding resident but instead will go extinct.

Figure 5 shows examples of such pairwise invasibility plots that were obtained using the same ecological functions as used in figures 1 and 3. In figure 5, regions in which the mutant can invade the corresponding resident are black, while regions in which the mutant cannot invade the corresponding resident are white. In both figure 5a (Gaussian ecological functions) and figure 5b (quartic ecological functions), the area below the diagonal is black, whereas the area above the diagonal is white (note that the diagonal itself belongs neither to the plus- nor to the minus-region, because a rare mutant with the same assortment phenotype as the resident will neither grow nor decline in total density, since the resident is at equilibrium). For very small resident values of $\sigma_A$, mutant growth rates are very close to zero. This is because in such resident populations any rare mutant has a strong effective assortment very similar to the resident, which is a consequence of the assumption that the probability of mating between two types is determined by the product of their respective preferences; see eq. (16) in the Appendix. Thus, for very small values of $\sigma_A$ selection as measured by initial mutant growth rates is nearly neutral, which is indicated by medium grey shading in figures 5a and 5b. Nevertheless, the figures show that there is directional selection for decreased $\sigma_A$, and hence for increased assortment. This is not surprising: selection favours increased assortment because assortative mating is a mechanism that facilitates the evolutionary response to frequency-dependent competition (Dieckmann & Doebeli 1999). This mitigation of frequency dependence manifests itself as pattern formation in phenotype space.

There are various ways in which assortative mating could incur fertility costs. One straightforward way to incorporate such costs in the models studied here is to assume that the intrinsic growth rate $r$ is negatively affected by increased assortment, i.e., by decreased $\sigma_A$. For example, we can replace the birth terms $\beta_i(x)$ in eqs. (13) and (14) by

$$\left[1 - c/(1+\sigma_{A_i})\right]\beta_i(x), \tag{15}$$

so that the new cost parameter $c$ determines the maximal fertility cost, incurred for very strong assortment (i.e., for $\sigma_A \to 0$). With costs of assortment, the pairwise invasibility plots change qualitatively, as is shown in figures 5c and 5d. For low resident values of $\sigma_A$, the plus- and minus-regions are now reversed across the diagonal, so that the plus-region is above the diagonal and the minus-region is below the diagonal. This means that for low resident values of $\sigma_A$ mutants with higher values of $\sigma_A$ than the resident, i.e., less assortative mutants, can invade, while more assortative mutants cannot. Thus, at low values of $\sigma_A$ there is directional selection for less assortative mating. However, at high values of $\sigma_A$ there is still directional selection for increased assortment (i.e., for lower $\sigma_A$). The point at which the two regimes of directional selection meet on the horizontal axis is an evolutionary attractor for the degree of assortment. Once the popula-tion has reached the corresponding degree of assortment, either from above or from below, no further invasion of nearby mutants occurs. As expected, costs of assortative mating thus move the evolutionary attractor for the trait $\sigma_A$ away from 0. Figures 5c and 5d show that even for moderately high costs of assortative mating, the degree of assortment is still expected to evolve to substantial levels.

Finally, we used an individual-based model to investigate the full evolutionary dynamics of assortment. In such a model, individuals are described by their ecological trait $x$ and by their assortment trait $\sigma_A$. At each point in time, every individual experiences a per capita death rate and a per capita birth rate. The per capita death rate is determined by the ecological trait and is calculated according to the death term in eq. (7) (integrals are replaced by sums over all individuals in the population). The per capita birth rate incorporates potential costs of assortment and is given by (15). At each point in time, individual rates are summed up to give the total birth and death rates $B$ and $D$, respectively. The waiting time until the next birth or death event is drawn from an exponential probability distribution with mean $1/(B+D)$, and a birth or death event is then chosen with probabilities $B/(B+D)$ and $D/(B+D)$, respectively. If a death event occurs, one individual is chosen probabilistically according to its relative contribution to the total death rate. The chosen individual is removed, and the birth and death rates of all other individuals are adjusted accordingly. If a birth event occurs, one individual is chosen probabilistically according to its relative contribution to the total birth rate. The chosen individual then selects a mating partner probabilistically according to the mate choice function given by eq. (16) in the Appendix, evaluated for all other individuals in the population (as before, mate choice is based on the ecological trait). The resulting mating pair produces an offspring whose phenotypes are drawn from two Gaussian distributions with means given by the midparent values of the two traits and with standard deviations $\sigma_f$ for the ecological trait and $\sigma_{f,\text{ass}}$ for the assortment trait. The offspring individual is inserted, and the birth and death rates of all other individuals are adjusted accordingly. This stochastic model naturally extends to finite populations the deterministic models introduced and analyzed above.

Figure 6 shows examples of the joint evolutionary dynamics of the ecological phenotype and the assortment phenotype in the individual-based model. The initial conditions for these dynamics were chosen such that populations were mating approximately randomly. As a consequence, the phenotype distribution for the ecological trait was initially unimodal (figure 6a). However, despite costs of assortment, assortative mating readily evolved to a degree that allowed the formation of phenotypic clusters, and hence diversification (figures 6b and 6c).

## 4. DISCUSSION

Our results show that even in infinitesimal genetic models for the dynamics of continuous phenotype distributions in sexual populations, frequency-dependent selection can split the population into separate phenotypic clusters when mating is assortative. If such pattern formation in phenotype space occurs, the emerging phenotypic clusters represent incipient species, as they are at least partially reproductively isolated due to assortative mating. Thus, frequency-dependent selection can cause adaptive speciation in these models.

Apart from assortative mating, two requirements need to be met for diversification to occur. The width of the offspring distribution produced by a given mating pair must be small enough, and frequency dependence must be strong enough. Our extensive numerical explorations of thousands of different cases revealed that when these conditions are satisfied, pattern formation occurs for a wide range of assortative mating parameters and for a wide range of forms of the competition kernel and the carrying capacity function. Moreover, assortative mating can readily evolve in initially randomly mating populations even if it comes at considerable cost. This is in accordance with recent results from explicit multilocus models showing that costs to assortment do not prevent adaptive sympatric speciation unless such costs are high (Doebeli & Dieckmann 2005; Doebeli 2005; Schneider & Bürger 2006; Bürger & Schneider 2006; Bürger et al. 2006). In this study, we have focused on assortative mating based on the ecological trait under frequency-dependent selection. Such assortment models are usually called one-allele models (Kirkpatrick & Ravigné 2002), in contrast with two-allele models, in which assortment is based on a selectively neutral display trait. The evolution of assortment in two-allele models for adaptive speciation has been studied in models with explicit multilocus genetics (Dieckmann & Doebeli 1999; Doebeli 2005), with the conclusion that, while recombination between the display trait and the ecological trait hinders adaptive diversification, speciation is nevertheless possible in such scenarios, even if there are costs to assortment (Doebeli 2005). It would clearly be interesting to study two-allele scenarios in infinitesimal models incorporating frequency-dependent competition.

Also, in this paper we have only considered costs to assortment that differentiate between different levels of assortative mating. For a population with a fixed degree of assortment, these costs are the same for all individuals. However, it is important to consider also scenarios in which there is a cost to assortment due to Allee effects. In this case, individuals may differ in fertility even in populations with a fixed degree of assortment, because rare phenotypes will encounter fewer preferred mates than common phenotypes, and so may have lower fertility. Allee effects can be incorporated into infinitesimal models (Noest 1997), and results for the dynamics of pattern formation in such models will be reported elsewhere. Implementing the individual-based model introduced at the end of the previous section is a straightforward exercise, and we invite readers to explore the dynamics of phenotypic pattern formation based on their own models and/or implementations. Whether the regions in parameter space for which diversification can be observed are biologically realistic is a question that needs to be addressed in empirical studies, but, mathematically speaking, it is clear that phenotypic pattern formation is a robust outcome of infinitesimal models.

In fact, diversification involving assortative mating may be easier in infinitesimal models than in other, more explicit genetic models based on a finite number of loci with finite effects, such as those investigated by Dieckmann & Doebeli (1999), Kirkpatrick & Nuismer (2004), Schneider & Bürger (2006), Bürger & Schneider (2006), and Bürger *et al.* (2006). One obstacle to speciation in such models is that genetic variation in the ecological trait can be exhausted if mating is strongly assortative (Kirkpatrick & Nuismer 2004; Bürger *et al.* 2006). This cannot happen in deterministic infinitesimal models, in which offspring distributions always range across the whole spectrum of phenotypes (albeit with very low frequencies at most phenotypes). In infinitesimal models all phenotypes are thus present at all times, and hence any loss of phenotypes on which selection can act is not a problem. At any rate, the results reported here for infinitesimal models are in surprisingly good overall agreement with models for adaptive diversification based on adaptive dynamics and on multilocus genetics (Dieckmann & Doebeli 1999; Doebeli & Dieckmann 2000; Bürger *et al.* 2006), supporting the understanding that adaptive speciation due to frequency-dependent interactions can safely be considered a theoretically plausible scenario.

It should, of course, be analyzed whether infinitesimal models can indeed provide a sufficiently accurate approximation of multilocus dynamics under frequency-dependent disruptive selection. This may well be the case over shorter time spans, involving mostly standing genetic variation (Bulmer 1980), but the robustness of this approximation becomes more uncertain when one takes into account mutation and substitution of allelic effects at the loci. It could happen that variation typically becomes concentrated on just one or a few loci (van Doorn & Dieckmann 2006), which is similar to the outcome predicted in adaptive dynamics models. Alternatively, variation might increase at all loci, potentially increasing $\sigma_f$, which could in principle result in a stable unimodal equilibrium distribution. In addition, there are other possible evolutionary responses to frequency-dependent competition, such as sexual dimorphism and a widening of individual niche widths (Bolnick & Doebeli 2003; Ackermann & Doebeli 2004; Rueffler *et al.* 2006) that could be considered. Although these are relevant issues, here we have focused our treatment on the infinitesimal model used by Polechová & Barton (2005) because, at the very least, it represents a conceptually interesting and traditionally well received case.

Our results are in contrast to those reported by Polechová & Barton (2005) for infinitesimal models in which both the competition kernel and the carrying capacity are Gaussian functions. Rather than focusing on the actual dynamics of phenotype distributions, results presented by Polechová & Barton (2005) only concern the variance of Gaussian equilibrium distributions: these authors seem to have implicitly assumed that the dynamics of the infinitesimal model always converges to such Gaussian solutions. In particular, Polechová & Barton (2005) did not consider the possibility that the Gaussian equilibrium could be unstable, nor did they investigate models with non-Gaussian ecological functions, which might not admit unimodal equilibrium solutions in the first place. Our results show that, in general, the dynamics of infinitesimal models do not converge towards unimodal equilibrium distributions when mating is assortative. While they do not appear to have numerically solved the infinitesimal model, Polechová & Barton (2005) mention that their simulations of a related model, the "symmetric" model with explicit multilocus genetics, suggest that dynamics in that symmetric model always converge to Gaussian equilibrium solutions, thus apparently lending support to their implicit assumption of stable Gaussian equilibrium distributions for the infinitesimal model. However, evaluating dynamical stability of one model in terms of another model is obviously not possible. Moreover, it has already been shown by Doebeli (1996) that the symmetric model also often exhibits pattern formation in the form of bimodal equilibrium distributions when mating is assortative.

Based on their analysis of the infinitesimal model, Polechová & Barton (2005) concluded that the process of assortment itself, irrespective of any frequency-dependent competition, was the most important driver of divergence in sexual models of sympatric speciation. This conclusion was based on observing that in the infinitesimal model with $\sigma_K = \infty$ and sufficiently strong assortment, the variance of a solution can increase without bound, even in the absence of frequency-dependent competition (i.e., if $\sigma_\alpha = \infty$). The possible role of assortment in permitting genetic divergence is of course a relevant issue. For the infinitesimal model, equilibrium solutions with infinite genetic variance exist. For instance, for an infinite width of the carrying capacity function, and with very strong assortative mating ($\sigma_A = 0$) but without frequency dependence, eq. (12) for determining the equilibrium variance $\sigma_{eq}$ reduces to $1/(\sigma_f^2 + \sigma_{eq}^2) = 1/\sigma_{eq}^2$, which only admits $\sigma_{eq} = \infty$ as a solution. This solution might be regarded as an artefact of the assumption of the infinitesimal model that there is an unlimited supply of genetic variation in the population. Nevertheless, the qualitative conclusion that assortment sometimes can relax genetic constraints and thus enable an increase in genetic variation seems valid.

An interesting question is then if assortment itself, without frequency-dependent competition, can lead to pattern formation. For our model, making the assumptions of very strong assortment ($\sigma_A = 0$), no frequency dependence ($\sigma_\alpha = y$), and a Gaussian carrying capacity with finite $\sigma_K$, we infer from eq. (12) that there is a Gaussian equilibrium distribution with finite variance $\sigma_{eq}^2 = \frac{1}{2}(-\sigma_f^2 + \sigma_f\sqrt{\sigma_f^2 + 4\sigma_K^2})$, which is approximately equal to $\sigma_f \sigma_K$ for small $\sigma_f$. Numerical simulations indicate that this unimodal solution is always stable. Similarly, when the carrying capacity is not of Gaussian form, but still unimodal, even very strong assortative mating never generated pattern formation in the absence of frequency dependence. Thus, in our numerical analysis of the infinitesimal model, speciation was never observed when frequency dependence was absent, independent of the strength of assortment.

We can gain an intuitive understanding of the reason for this conclusion by noting that eq. (7) in the limit of $\sigma_A \to 0$ approaches the asexual case in eq. (2), except that the offspring distribution $N_{(u+v)/2,\sigma_f}$ acts like a mutation kernel, smoothing the distribution $\phi$. For such an asexual model, with very high rate of mutation and no frequency-dependent competition, there is no reason to expect clustering of phenotypes, since there are no forces that could counteract the homogenization of a multimodally clustered distribution. Thus, under very strong assortment the sexual production of offspring, involving segregation and recombination, can increase genetic variation in the infinitesimal model, in a manner analogous to the process of mutation. This source of variation, however, cannot in itself drive pattern formation, just as mutation cannot in itself drive pattern formation. In the infinitesimal models considered here and in Polechová & Barton (2005) speciation is thus impossible without frequency-dependent competition.

The results reported here are in complete agreement with those of Noest (1997), who presented an analytical study of a special class of infinitesimal models, in which the carrying capacity was assumed to be uniform (i.e., independent of $x$), and the competition kernel was assumed to be Gaussian. These models admit a uniform equilibrium phenotype distribution, and Noest (1997) investigated the conditions under which this uniform solution becomes unstable in the presence of assortative mating. His analytical results match our numerical results in essential aspects: if the offspring distribution is sufficiently narrow and frequency dependence is sufficiently strong, then the uniform solution can become unstable when mating is assortative. Noest (1997) did not study more general forms of the ecological functions, for which analytical results are not feasible, and he did not consider the evolution of assortative mating. However, his results already clearly showed that frequency dependence and assortative mating can break up the sexual continuum (Maynard Smith & Szathmáry 1995) through the formation of multimodal distributions in phenotype space. Our results lead to the same conclusion for a more general class of models and evolutionary scenarios: adaptive speciation can occur as a result of pattern formation in phenotype space due to frequency-dependent selection and the evolution of assortative mating.

**Acknowledgements:** MD was supported by NSERC (Canada) and by the James S. McDonnell Foundation (USA). OL was supported by the Swedish Research Council. UD acknowledges financial support by the Vienna Science and Technology Fund, WWTF.

**REFERENCES**

Ackermann, M. & Doebeli, M. 2004 Evolution of niche width and adaptive diversification. *Evolution* **58**, 2599–2612.

Bolnick, D. I. & Doebeli, M. 2003 Sexual dimorphism and adaptive speciation: two sides of the same ecological coin. *Evolution* **57**, 2433–2449.

Bolnick, D. I., Svanbäck, R., Fordyce, J. A., Yang, L. H., Davis, J. M., Hulsey, C. D., & Forister, M. L. 2003 The ecology of individuals: Incidence and implications of individual specialization. *Am. Nat.* **161**, 1–28.

Bulmer, M. G. 1980. *The mathematical theory of quantitative genetics.* Oxford, UK: Clarendon Press.

Bürger, R. & Schneider, K. A. 2006 Intraspecific competitive divergence and convergence under assortative mating. *Am. Nat.* **167**, 190–205.

Bürger, R., Schneider, K. A., & Willensdorfer, M. 2006 On the conditions for speciation through intraspecific competition. *Evolution*, in press.

Coyne, J. A. & Orr, H. A. 2004 *Speciation.* Sunderland, MA: Sinauer Associates.

Dieckmann, U. & Doebeli, M. 1999 On the origin of species by sympatric speciation. *Nature* **400**, 354–357.

Dieckmann, U., Doebeli, M., Metz, J. A. J., & Tautz, D. (eds.) 2004 *Adaptive speciation.* Cambridge, UK: Cambridge University Press.

Doebeli, M. 1996 A quantitative genetic competition model for sympatric speciation. *J. Evol. Biol.* **9**, 893–909.

Doebeli, M. 2005 Adaptive speciation when assortative mating is based on female preference for male marker traits. *J. Evol. Biol.* 18, 1587–1600.

Doebeli, M. & Dieckmann, U. 2000 Evolutionary branching and sympatric speciation caused by different types of ecological interactions. *Am. Nat.* **156**, S77–S101.

Doebeli, M. & Dieckmann, U. 2003 Speciation along environmental gradients. *Nature* **421**, 259–264.

Doebeli, M. & Dieckmann, U. 2005 Adaptive dynamics as a mathematical tool for studying the ecology of speciation processes. *J. Evol. Biol.* **18**, 1194–1200.

Drossel, B. & McKane, A. 2000 Competitive speciation in quantitative genetic models. *J. Theor. Biol.* **204**, 467–478.

Ellner, S. & Sasaki, E. 1995 The evolutionarily stable phenotype distribution in a random environment. *Evolution* **49**, 337–350.

Geritz, S. A. H., Kisdi, É., Meszéna, G., & Metz, J. A. J. 1998 Evolutionarily singular strategies and the adaptive growth and branching of the evolutionary tree. *Evol. Ecol.* **12**, 35–57.

Geritz, S. A. H. & Kisdi, É. 2000 Adaptive dynamics in diploid, sexual populations and the evolution of reproductive isolation. *Proc. R. Soc. B* **267**, 1671–1678.

Kirkpatrick, M. & Nuismer, S. L. 2004 Sexual selection can constrain sympatric speciation. *Proc. R. Soc. B* **271**, 687–693.

Kirkpatrick, M. & Ravigné, V. 2002 Speciation by natural and sexual selection: models and experiments. *Am. Nat.* **159**, S22–S35.

Kisdi, É. & Geritz, S. A. H. 1999 Adaptive dynamics in allele space: evolution of genetic polymorphism by small mutations in a heterogeneous environment. *Evolution* **53**, 993–1008.

Kisdi, É. & Gyllenberg, M. 2005 Adaptive dynamics and the paradigm of diversity. *J. Evol. Biol.* **18**, 1170–1173.

Kondrashov, A. S. & Kondrashov, F. A. 1999 Interactions among quantitative traits in the course of sympatric speciation. *Nature* **400**, 351–354.

MacArthur, R. 1972. *Geographical ecology.* New York: Harper and Row.

Maynard Smith, J. & Szathmáry, E. 1995. *The major transitions in evolution.* Oxford, UK: W.H. Freeman.

Meszéna, G., Gyllenberg, M., Jacobs, F. J. A., & Metz, J. A. J. 2005 Link between population dynamics and dynamics of Darwinian evolution. *Phys. Rev. Lett.* 95, Art. No. 078105.

Metz, J. A. J., Geritz, S. A. H., Meszéna, G., Jacobs, F. J. A., & van Heerwaarden, J. 1996. Adaptive dynamics: a geometrical study of the consequences of nearly faithful reproduction. In *Stochastic and spatial structures of dynamical systems.* (S. van Strien & S. Verduyn Lunel eds.) pp. 183–231 Dordrecht, Netherlands: North Holland.

Metz, J. A. J., Nisbet, R. M., & Geritz, S. A. H. 1992 How should we define 'fitness' for general ecological scenarios? *Trends Ecol. Evol.* **7**, 198–202.

Noest, A. J. 1997 Instability of the sexual continuum. *Proc. R. Soc. B* **264**, 1389–1393.

Polechová, J. & Barton, N. H. 2005 Speciation through competition: a critical review. *Evolution* **59**, 1194–1210.

Roughgarden, J. 1979. *Theory of population genetics and evolutionary ecology.* New York: Macmillan.

Rueffler, C., Van Dooren, T. J. M., Leimar, O., & Abrams, P. 2006 Disruptive selection and then what? *Trends Ecol. Evol.* **21**, 238–245.

Sasaki, A. 1997 Clumped distribution by neighborhood competition. *J. Theor. Biol.* **186**, 304–329.

Schneider, K. A. & Bürger, R. 2006 Does competitive divergence occur if assortative mating is costly? *J. Evol. Biol.* **19**, 570–588.

van Doorn, G. S. & Dieckmann, U. 2006 The long-term evolution of multi-locus traits under frequency-dependent disruptive selection. *Evolution*, in press.

## APPENDIX

Here we derive expressions for the birth rates $\beta_1(x)$ and $\beta_2(x)$ used in eqs. (13) and (14). The probability of mating between type 1 and type 2 is the product of their respective preferences determined by $\sigma_{A_1}$ and $\sigma_{A_2}$. Thus, for a given phenotype $u$ of mating type 1, the probability of mating with a phenotype $v$ of mating type $i$ (where $i=1$ or $i=2$) is proportional to

$$A_{1i}(u,v) = \frac{1}{\sqrt{2\pi}\sigma_{A_1}} \exp\left(-\frac{(u-v)^2}{2\sigma_{A_1}^2}\right) \cdot \frac{1}{\sqrt{2\pi}\sigma_{A_i}} \exp\left(-\frac{(u-v)^2}{2\sigma_{A_i}^2}\right). \tag{16}$$

In accordance with eq. (7) for the single-type case, the offspring distribution of type 1 is then given by

$$\beta_1(x) = \int \phi_1(u) \left[ \frac{1}{N_1(u)} \int [\phi_1(v) A_{11}(u,v) + \phi_2(v) A_{12}(u,v)] N_{(u+v)/2,\sigma_f}(x) dv \right] du, \tag{17}$$

where

$$N_1(u) = \int [\phi_1(v) A_{11}(u,v) + \phi_2(v) A_{12}(u,v)] dv \tag{18}$$

provides the normalization necessary to ensure that, up to explicit costs of assortment, all phenotypes have the same total reproductive output. An analogous formula holds for $\beta_2(x)$.

If there is only one assortment type present, with assortative mating parameter $\sigma_{A_1}$, the above two-type model reverts to the original single-type model.

**FIGURE LEGENDS**

**Figure 1.** Equilibrium phenotype distributions for different degrees of assortative mating with Gaussian competition kernel and Gaussian carrying capacity function, corresponding to $\varepsilon_\alpha = \varepsilon_K = 0$ in eqs. (8) and (9). (*a*) Random mating ($\sigma_A = \infty$) does not allow pattern formation. (*b,c*) Assortative mating (with $\sigma_A = 0.56$ and $\sigma_A = 0.28$, respectively) can generate multimodal phenotype distributions. (*d*) With Gaussian ecological functions, very strong assortative mating ($\sigma_A = 0$) leads to unimodal phenotype distributions. In each panel, the grey curve shows the carrying capacity function $K$. Other parameters: $r = 1$, $K_0 = 1$, $\sigma_K = 2$, $\sigma_\alpha = 1$, and $\sigma_f = 0.2$; initial phenotypes distribution were Gaussian with variance equal to that of a solution of eq. (10); dynamics were run for $10^4$ time units.

**Figure 2.** Dynamics of the system shown in figure 1b for different initial conditions (shown as thick curves). (*a*) Gaussian initial distribution with variance equal to $\sigma_A^2$. (*b*) Gaussian initial distribution with variance equal to that of a numerical solution of eq. (10). (*c*) Uniform initial distribution (i.e., the phenotypic density is independent of $x$). In each case, the total density of the initial distribution was equal to the total density of the Gaussian equilibrium distribution whose variance is given by eq. (10). In all cases, phenotype distributions converge to the trimodal equilibrium shown in figure 1b.

**Figure 3.** Equilibrium phenotype distributions for different degrees of assortative mating with quartic competition kernel and carrying capacity function, corresponding to $\varepsilon_\alpha = \varepsilon_K = 2$ in eqs. (8) and (9). (*a*) Random mating ($\sigma_A = \infty$) does not allow pattern formation. (*b-d*) When assortative mating is strong enough ($\sigma_A = 0.56$, $0.28$, and $0$ in (*b*) to (*d*), respectively), it always allows multimodal phenotype distributions. In each panel, the grey curve shows the carrying capacity function $K$. Other parameters: $r = 1$, $K_0 = 1$, $\sigma_K = 2$, $\sigma_\alpha = 1$, and $\sigma_f = 0.2$; initial phenotypes distribution were Gaussian with variance 1; dynamics was run for $10^4$ time units.

**Figure 4.** Dependence of the number of modes in equilibrium phenotype distributions on ecological functions and assortative mating. (*a*) Gaussian competition kernel. Deviations from Gaussian form in the carrying capacity function are measured by $\varepsilon_K$ in eq. (9), ranging from $-0.8$ to $0.8$ in increments of $0.05$ along the vertical axis. The degree of assortment is measured by $\sigma_A$ in eq. (3), ranging from $0$ to $1.41$ in increments of $0.07$ along the horizontal axis. For each grid point, the infinitesimal model in eq. (7) was run from uniform initial conditions to equilibrium ($10^4$ time units). Grey levels indicate whether the equilibrium distribution was unimodal (dark grey) or multimodal (light grey); intermediate grey levels indicate interpolated regions in which modality is uncertain. (*b*) Gaussian carrying capacity function. Deviations from Gaussian

form in the competition kernel are measured by $\varepsilon_\alpha$ in eq. (8), ranging from $-0.8$ to $0.8$ in increments of $0.05$ along the vertical axis. Other settings as in (*a*). Other parameters: $r = 1$, $K_0 = 1$, $\sigma_K = 2$, $\sigma_\alpha = 1$, and $\sigma_f = 0.2$.

**Figure 5.** Pairwise invasibility plots for the degree of assortment. In each plot, the horizontal axis shows resident values of $\sigma_A$ and the vertical axis mutant values of $\sigma_A$. Regions are shaded according to whether for resident-mutant pairs in that region the rare mutant's growth rate is positive (black) or negative (white). Intermediate grey levels indicate regions in which the rare mutant's growth rate did not differ from 0 by more than $10^{-5}$. (*a*) Gaussian competition kernel and Gaussian carrying capacity function, corresponding to $\varepsilon_\alpha = \varepsilon_K = 0$ in eqs. (8) and (9). (*b*) Quartic competition kernel and carrying capacity function, corresponding to $\varepsilon_\alpha = \varepsilon_K = 2$ in eqs. (8) and (9). Without costs of assortment, the pairwise invasibility plots show directional selection for increased assortative mating. (*c,d*) Same as (*a,b*), but with the cost of assortment set to $c = 0.75$ in (15). Now there is an intermediate level of assortment to which directional evolution converges from both above and below. The corresponding intermediate resident degree of assortment cannot be invaded by any mutants (as indicated by the white regions) and hence is evolutionarily stable. Other parameters: $r = 1$, $K_0 = 1$, $\sigma_K = 2$, $\sigma_\alpha = 1$, and $\sigma_f = 0.2$. The pairwise invasibility plots were obtained by varying resident and mutant values of $\sigma_A$ from 0 to 1 in increments of $0.0375$. For each resident-mutant pair, the resident's phenotype distribution was first allowed to equilibrate from a flat initial distribution for $10^4$ time units, before a rare mutant with the same distribution shape as the resident but low total density was introduced; the mutant's growth rate was then measured over 100 time units.

**Figure 6.** Evolutionary dynamics of the individual-based model for Gaussian and quartic ecological functions and for costs of assortment set to $c = 1$ in (15). (*a*) In both cases, the initial distribution in the two-dimensional phenotype space was chosen to describe a randomly mating population (high values of $\sigma_A$) situated at the maximum of the carrying capacity function. (*b*) Evolutionary outcome for Gaussian ecological functions. (*c*) Evolutionary outcome for quartic ecological functions. After $10^4$ time units, phenotype distributions in both cases have moved into the region in which mating is assortative, permitting multimodality in the ecological phenotype. The shown bimodal distributions are stable and no longer change appreciably. This illustrates that pattern formation through the evolution of assortative mating is possible even if assortative mating is costly and the degree of assortment is inherited sexually. Other parameter values were the same as in figure 5, except for $K_0$ in eq. (9), which was set to $K_0 = 600$ (in the individual-based model, this parameter can be used to scale the total population size, which equalled approximately 500 individuals in the shown simulations). The parameter $\sigma_{f,\text{ass}}$ describing the width of the offspring distribution in the direction of the assortment trait was set to $0.05$.

Figure 1

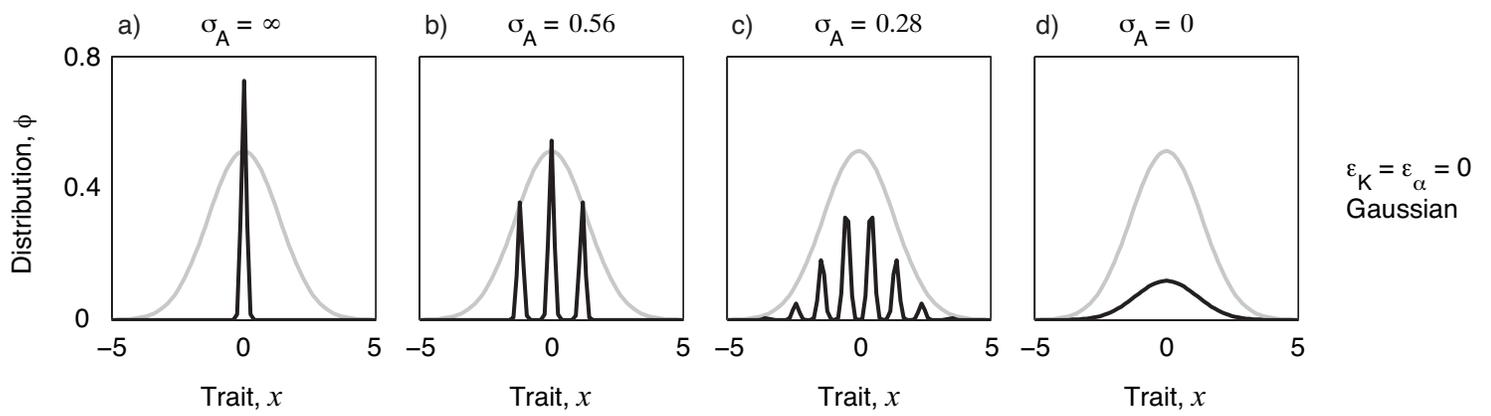

Figure 2

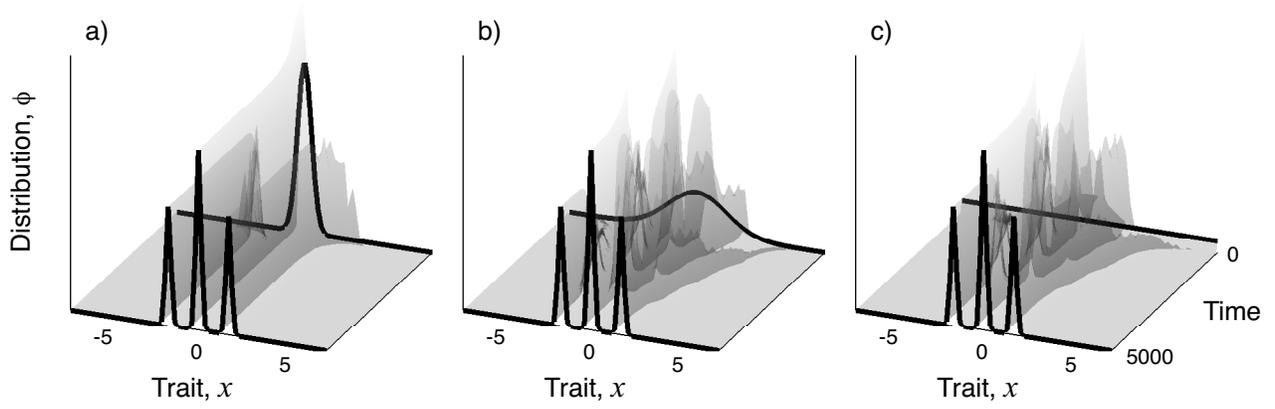

Figure 3

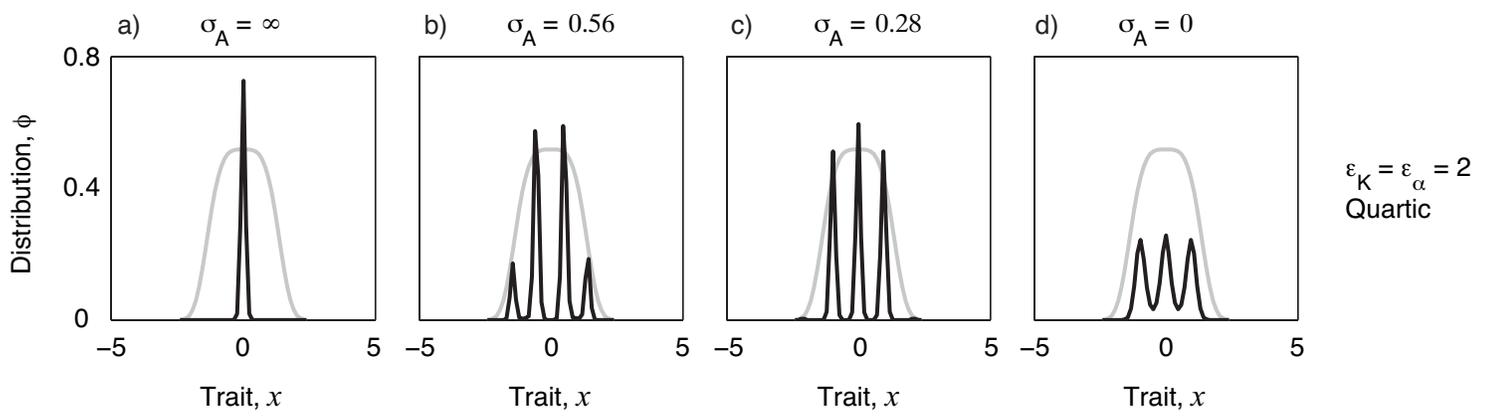

Figure 4

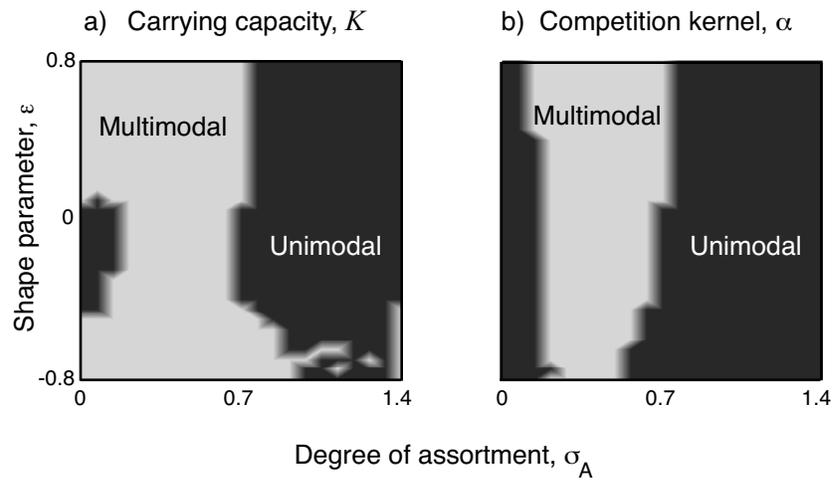

Degree of assortment, $\sigma_A$

Figure 5

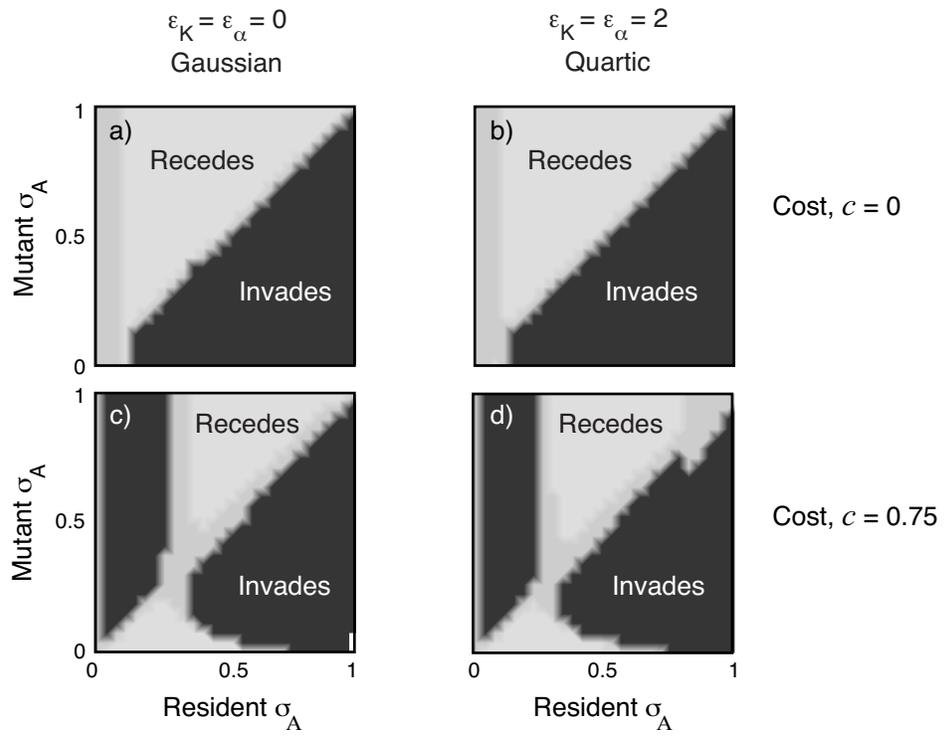

Figure 6

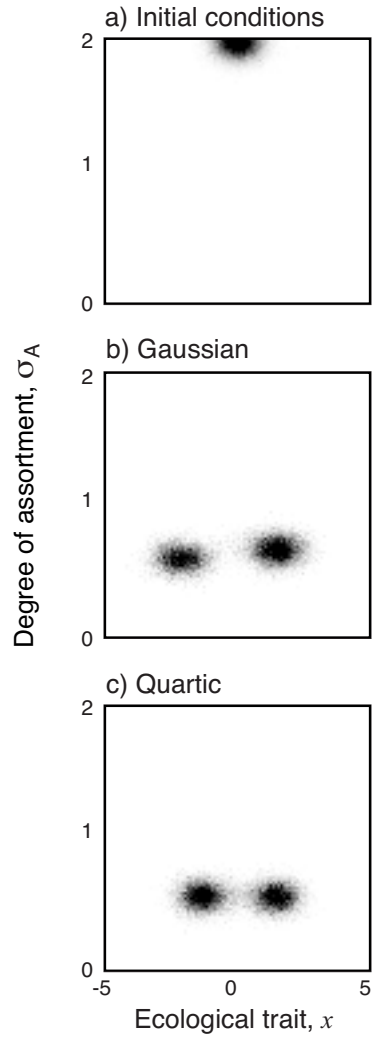